# Novel method distinguishing between competing topological orders

Bivas Dutta, Wenmin Yang, Ron Aharon Melcer, Hemanta Kumar Kundu, Moty Heiblum[*], Vladimir Umansky, Yuval Oreg, Ady Stern and David Mross

*Broun Center for Sub-Micron Research, Dep. of Condensed Matter Physics,*

*Weizmann Institute of Science, Rehovot, Israel 76100*

*e-mail: moty.heiblum@weizmann.ac.il

**Quantum Hall states – the progenitors of the growing family of topological insulators -- are rich source of exotic quantum phases. The nature of these states is reflected in the gapless edge modes, which in turn can be classified as integer - carrying electrons, fractional - carrying fractional charges; and neutral – carrying excitations with zero net charge but a well-defined amount of heat. The latter two may obey anyonic statistics, which can be abelian or non-abelian. The most-studied putative non-abelian state is the spin-polarized filling factor $\nu$ =5/2, whose charge $e$/4 quasiparticles are accompanied by neutral modes. This filling, however, permits different possible topological orders, which can be abelian or non-abelian. While numerical calculations favor the non-abelian anti-Pfaffian (A-Pf) order to have the lowest energy, recent thermal conductance measurements suggested the experimentally realized order to be the particle-hole Pfaffian (PH-Pf) order. It has been suggested that lack of thermal equilibration among the different edge modes of the A-Pf order can account for this discrepancy. The identification of the topological order is crucial for the interpretation of braiding (interference) operations, better understanding of the thermal equilibration process, and the reliability of the numerical studies. We developed a new method that helps identifying the topological order of the $\nu$ =5/2 state. By creating an interface between the two 2D half-planes, one hosting the $\nu$ =5/2 state and the other an integer $\nu$ =3 state, the interface supported a fractional $\nu$ =1/2 charge mode with ½ quantum conductance and a neutral Majorana mode. The presence of the Majorana mode, probed by measuring noise, propagating in the opposite direction to the charge mode, asserted the presence of the PH-Pf order but not that of the A-Pf order.**

The quantum Hall effect (QHE), harboring an insulating bulk and conductive edges, is the earliest known example of topological insulators[1]. It is characterized by topological invariants, which are stable to small changes in the details of the system[2]. Of these quantities, the easiest to probe is the electrical Hall conductance of the edge mode, $G_H = \nu e^2/h$, where $\nu$ is the bulk filling factor (integer or fractional), $e$ the electron charge, and $h$ the Planck constant. Quantum Hall states with fractional filling factors support quasiparticles with fractional charges and anyonic statistics. The ubiquitous Laughlin states are abelian, and an exchange of their quasiparticles positions adds a phase to the ground-state wave function[3-5]. In their more exotic cousins, the non-abelian ones[6,7], the presence of certain quasiparticles results in multiple degenerate ground states, and interchanging these quasiparticles cycles the system between different ground states. In general, QH states permit different topological orders, and the ubiquitous conductance and charge measurements are not sufficient to distinguish between the different topological orders.

Here, a second topological invariant comes into play, which is sensitive to all energy-carrying edge modes, charged or neutral: the thermal QH conductance, $G_T = KT$, with $K$ the thermal conductance coefficient and $T$ the temperature. It was proposed[7,8] and experimentally proven, that $K$ of a single chiral and ballistic mode, fermionic[9], bosonic[10], or (abelian) anyonic[11], is quantized $K = \kappa_0$, with $\kappa_0 = \pi^2 k_B^2/3h$, where $k_B$ is the Boltzmann's constant. However, a fractional value of $K$ is expected for non-abelian states[12]. Indeed, for the $\nu = 5/2$ state, we found a thermal Hall conductance coefficient, $K = 2.5 \kappa_0$ - consistent with the non-abelian PH-Pf topological order[13].

However, there is a caveat: For a QH state that supports multiple edge modes, some moving *downstream* (DS, in the chirality dictated by the magnetic field), and some *upstream* (US, in opposite chirality), the theoretically predicted thermal conductance assumes a full thermal equilibration among all modes[14-16]. For modes of integer thermal conductance, for example, $G_T = (n_d - n_u)\kappa_0 T$, where $n_d$ ($n_u$) are the number of DS (US) modes. In the other extreme, with no thermal equilibration, $G_T = (n_d + n_u)\kappa_0 T$ [11]. It is thus obvious, that the thermal equilibration length, being in general longer than the charge equilibration length, is of crucial importance in interpreting thermal conductance measurements.

Moore and Read were the first to predict that the topological order of the $\nu = 5/2$ state in a Pfaffian (Pf) state, with $K = 3.5 \kappa_0$ [17]. Yet, since an US neutral mode was found earlier in the $\nu = 5/2$ state[18], the Pf order, which does not support an US mode, was ruled out. Subsequent numerical studies that include Landau-level mixing favored the non-abelian particle-hole conjugate of the Pf order, the anti-Pf (A-Pf) with $K = 1.5 \kappa_0$, to be energetically favorable in a homogenous system (neglecting disorder)[19-24]. Several theoretical proposals offer possible explanations for the discrepancy between numerical calculations and the experimentally found PH-Pf order: *i.* Disorder may lead to islands of local Pf and A-Pf orders, from which a global PH-Pf order emerges[25-28]; ***ii.*** A considerably longer thermal equilibration length than the size of the device, may lead to deviation from the expected theoretical value. In particular, an unequilibrated Majorana mode in the A-Pf[14,29] order will add its contribution to *K* instead of subtracting it, leading to *K*=2.5 $\kappa_0$ [30-33]; ***iii.*** The numerical calculations, which predict an A-Pf order, do not take into account the unavoidable disorder, which may stabilize the PH-Pf order[29].

To make a definite determination of the topological order, one may suggest measuring the thermal conductance at a short propagation length, where thermal equilibration is practically negligible, expecting $K = (n_d + n_u)\kappa_0$; namely, $K_{A-Pf}$=4.5 $\kappa_0$ or $K_{PH-Pf}$=3.5 $\kappa_0$. However, 'spontaneous edge reconstruction' may add short-lived pairs of counter-propagating (non-topological) neutral modes[30], thus increasing the apparent thermal conductance of the states at short distances.

Our new experimental approach is to probe the chirality of the Majorana neutral modes at an engineered interface between $v$=5/2 state and $v$=2 & $v$=3 integers. It fulfills two important requirements: ***i.*** Thermally non-equilibrium: measurements are performed in the *intermediate* length regime; namely, longer than the charge equilibration length (leading to the quantized interface conductance), but shorter than the thermal equilibration length (allowing unequilibrated modes transport)[14,31,32]; ***ii.*** Gapping integer modes: the integer modes, being a part of $v$=5/2 state are fully gapped out at the interface, and thus flow only around the physical edge of the mesa. This eliminates the spurious hotspots and allows the identification of the topological order.

We employed a gated high-quality GaAs-AlGaAs MBE grown heterostructures (ED Sec. A). Structures were designed to resolve the contradictory requirements of the doped layers, which should ensure a full quantization of the fragile $v$=5/2 state, and, at the same time, allow highly stable and 'hysteresis free' operation of the gated structures[13]. Two separate gates divided the surface into two, an upper half and lower half (Fig. 1a). The gates were isolated from the sample and from each other by 15-25nm layers of $HfO_2$ (for more details of the structure see ED Sec. A & B). A gate-voltage in the range -1.5V<$V_g$<+0.3V allows varying the electron density from pinch off to 3x10$^{11}$cm$^{-2}$ (Fig. ED1). Tuning each gate separately controls the two interfaced filling factors, leading to the desired interface modes' conductance. The ohmic-contacts at the physical edge of the mesa probe the filling factors of the two respective half-planes, while the contacts at the interface probe the interface.

The heart of the measurement setup (highlighted by dashed-box in Fig. 1a) consists of three ohmic contacts at the interface, with the Source, S, placed symmetrically between the Amplifier contact, A, and the cold-grounded Drain, D. An injected DC Source current at the interface, $I_S$, forms a hotspot at the backside of the Source (and at the front-side of drain)[33]. In states that support counter-propagating modes, the thermally activated modes by the hotspot (usually, neutral modes), lead to shot noise at the amplifier contact[34,35]. The noise is filtered by an LC resonance circuit with a center frequency $f_0$~630kHz and bandwidth $\Delta f$=10-30kHz; amplified by a cold amplifier (cooled to 4.2K) and a room temperature amplifier, to be measured by a spectrum analyzer. Measurements are conducted at three different electron temperatures, 10mK, 21mK, and 28mK, and at different propagation lengths (between Source and Amplifier contacts), 28μm, 38μm, 48μm, and 58μm.

The general strategy of the measurement is to place the lower-gated half-plane to the *tested* filling factor $\nu_{LG}$, while upper half-plane is always tuned to an integer filling, $\nu_{UG}$=0, 1, 2, 3. The defined chirality of the resultant interface edge modes is always with respect to the chirality of the tested state. For example, for $\nu_{LG} > \nu_{UG}$ the interface charge chirality is DS, while for $\nu_{LG} < \nu_{UG}$ the interface charge chirality is US. This is clearly noted in all the figures (see also ED Sec. C). In order to measure the noise excited by the hotspot (at the back of the Source) by a single amplifier, the magnetic field was reversed (and so the chirality) under these two interfacing conditions. In the figures, for convenience, we flip the amplifier's position instead of the chirality (e.g. Figs. 1b & 1c).

In Figs. 1b & 1c, we show a relatively simple experimental test of interfacing the integer $\nu$=2 state (being the tested state) with $\nu$=1 and $\nu$=3. The injected DC current, $I_S$, leads to a hotspot at the US side of the Source in Fig. 1b and at the DS side of the Source in Fig. 1c, respectively. A perfect charge equilibration took place for all four lengths and three temperatures, with $R_S$=$h/e^2$; without an observed US (Fig. 1b)/DS (Fig. 1c) noise. This observation suggests that the presence of any residual non-equilibrated current (which may persist despite of charge equilibration) does not lead to any observable noise.

Before testing interfacing in the second LL, we test the 'interfacing method' with more complex abelian state in the first LL, which involves counter-propagating modes (charge and neutral). We interface the $\nu$=5/3=1+2/3 filling with the integers $\nu$=0, 1, 2. Two similar descriptions of this configuration are employed (see Fig. ED3). The first assumes that the two half-plans are initially separated, and each of them supports its own edge modes (Figs. 1e &1f). With an intimate proximity at the interface, the integer modes on both sides of the interface gap each other, leaving only fractional interface modes. In the case of 5/3 interfaces, the integer modes of $\nu$=1 or $\nu$=2 are gapped, with interface DS $\nu$=2/3 (with neutral) or US $\nu$=1/3 modes, respectively (Fig. ED 3e, 3g). The second approach is to regard the integer filling $\nu_{UG}$ at the upper half-plane as a "vacuum" on which a filling $\nu_{LG}$-$\nu_{UG}$ resides in the lower half-plane. Consequently, the common $\nu_{UG}$ integer modes circulate around the periphery of the mesa, and the interface carries an edge structure of filling $\nu_{LG}$-$\nu_{UG}$ (see Figs. ED 3f & 3h). We mostly use the first approach.

In Figs 1d-1f we return to the present test of interfacing the 5/3 state. Interfacing 5/3-0 (and 5/3-1), support an integer and a fractional $\nu$=2/3 charge modes (a fractional $\nu$=2/3 charge mode), accompanied by an excited US bosonic neutral mode[36-38], leading to the observed US noise (Fig. 1d & 1e). Alternatively, interfacing 5/3-2, the gapped integers leave behind a single US $\nu$=1/3 mode at the interface, with no noise observed (Fig. 1f).

We concentrate now on interfacing the dominant fractional states in the second Landau level, $\nu$=7/3, $\nu$=5/2, and $\nu$=8/3 with the integers $\nu$=2 and $\nu$=3. Testing first charge equilibration, we fixed the integer filling in the upper half-plane and swept the density of the lower half-plane (Fig. 2a). Clear conductance plateaus, accurate to about 1% (with re-entrant peaks and valleys between plateaus), are observed at all propagation lengths and temperatures.

The interfaced 7/3-2 configuration gaps the two integer modes leaving a DS edge mode of $\nu$=1/3, with no US noise (see Fig. 2b; the chirality is drawn in Fig. 2a). In contrast, the interfaced 7/3-3 configuration leaves the familiar US $\nu$=2/3 charge mode and a DS bosonic neutral mode (Fig. 2c). The hotspot at the Source excites the neutral mode with a DS noise at the amplifier. Interfacing $\nu$=8/3 with $\nu$=2 gaps the two integers and leaves a DS charge mode of $\nu$=2/3 and an US excited bosonic neutral mode (Fig. 2d). However, interfacing the state with $\nu$=3 leaves an US $\nu$=1/3 charge mode at the interface, with no resulting noise (Fig. 2e).

Before showing the main experimental results, it is worth discussing the outcome of interfacing the PH-Pf and A-Pf orders of ν=5/2 with the integers ν=0, 2, 3 (Fig 3a-3f). The consequence of a similar interfacing of other proposed orders of ν=5/2 are described in ED Sec. G. The mode structure of the two topological orders of ν=5/2 with vacuum, i.e. at the 5/2-0 interface, are shown in Figs. 3a & 3b. The 5/2-2 interface leaves for both the orders, a DS fractional charge mode $\nu$=1/2 and US Majorana modes; one for the PH-Pf and three for A-Pf orders (Figs. 3c & 3d). The 5/2-3 interface is more interesting. Interface of the PH-Pf ν=5/2 with ν=3, supports counter-propagating US fractional charge mode $\nu$=1/2 and a DS Majorana mode (Fig. 3e); while for the interface of the A-Pf ν=5/2 and ν=3, the latter two modes co-propagate in the US direction (Fig. 3f). Therefore, measuring the chirality of the Majorana mode at the 5/2-3 interface is crucial for identifying the actual topological order of the ν=5/2 state.

Figures 3g & 3h show the noise data for 5/2 interfaces, measured at 10mK with the propagation length of 28μm. Noise was found in the US direction at the interface 5/2-2 (no noise observed in DS), and in the DS direction for 5/2-3 (no US noise was observed) - both with similar amplitude at the same Source voltage. As discussed above, the measured DS noise at the 5/2-3 interface points at the existence of the PH-Pf order (Fig. 3h, inset). Measurements in all temperatures and lengths (with two different MBE growths and two thermal cycles) led to similar results (see ED Sec. I). This is the main result of our work.

It is worth noting that the PH-Pf is a particle-hole symmetric state. Hence, the same outcome should occur when it is interfaced with $\nu$ =2, where the system is regarded as a half-filled level of electrons on top of two full Landau levels; and with $\nu$ =3, where the system is regarded as a half-filled Landau level of holes on top of three full Landau levels. Our results indeed manifest this particle-hole symmetry.

The amplitude of the neutral noise as function of the number of gapped integer modes is also important. Since the temperature of the hotspot is proportional to the applied voltage ($T^2_{HS} \sim KV_S^2$), we plotted the noise data as a function of the Source voltage at a fixed propagation length of 28μm, for a few interfacing conditions (Fig. 4a). The noise (US or DS) was similar for all integers $n$. The same behavior was also observed for the 8/3-$n$ interfaces (see Fig. ED 16b). These results indicate that the integers' hotspots, located at the boundary of the large Source ohmic contact, do not take part in the excitation of the neutral (bosonic or Majorana) modes. The latter are excited by the hotspot generated by inter-edge equilibrations, somewhat remote from the Source contact[14,31]. The dependence of the noise as function of the propagation length, for the 5/2-$n$ and 8/3-$n$ interfaces, is shown in Fig. 4b (the solid lines are qualitative fits drawn as a guide to the eye), indicating a qualitatively similar thermal equilibration process of the 'different neutrals' as the propagation length is increased. In ED Sec. J, we add similar measurements of the interfaces 5/3-$n$ and 2/3-0.

As we showed above the measured results are quite naturally explained in terms of interfaces between a PH-Pf topological order of ν=5/2 and the integers $\nu$ =2 and $\nu$ =3. We however cannot exclude the possibility of edge reconstruction that might give rise to noise at the interface. In particular, this may include a scenario in which the presence of an interface to an integer filling makes the topological order of the $\nu = 5/2$ close to the interface different from that of the bulk. A specific example of an interface reconstruction in the A-Pf bulk is discussed in ED Sec. H. We note, however, that there is no indication for such edge reconstruction in the whole set of the measurements we performed on interfaces of the Abelian cases.

In this work we have introduced a novel method that is instrumental in identifying the topological order of the non-abelian $\nu$ =5/2 state. Since the previous experimental determination of the PH-Pf order was based on full thermal equilibration of all modes[11,13], questions were raised whether this condition was fulfilled[39]. Here, by forming *chiral 1D modes* at the interface between two half-planes, each with a different filling factor, leading to a single $\nu$ =1/2 charge mode and Majorana modes, allowing to assert the presence of a PH-Pf topological order. Our measurements also rule out all other, a-priory possible, topological orders of the $\nu$ =5/2[6,40-42] (ED Sec. G). In a broader perspective, introducing a similar 'interfacing method' between quantum states may also bring to life new chiral 1D modes, which do not live on the edge. In the present circumstances, it opens up a new test bench that can be applied to other unexplored exotic states.

**Acknowledgments**

We acknowledge valuable discussions with B. Halperin, M. Banerjee and N. Schiller. B.D. acknowledges the support from Clore Foundation, D. Mahalu for her help with E-beam-lithography, R. Bhattacharya for technical help, M.H. acknowledges the continuous support of the Sub-Micron Center staff. M.H. acknowledges the support of the European Research Council under the European Community's Seventh Framework Program (FP7/2007-2013)/ERC under grant agreement number 713351, the partial support of the Minerva foundation under grant number 713534, and together with V.U. the German Israeli Foundation (GIF) under grant number I-1241-303.10/2014. D.M acknowledges support from the ISF (1866/17) and the CRC/Transregio 183. Y.O. and A.S. acknowledge partial support through the ERC under the European Union's Horizon 2020 research and innovation program (grant agreement LEGOTOP No 788715), the ISF Quantum Science and Technology (2074/19), and the CRC/Transregio 183. YO acknowledges support from the BSF and NSF (2018643).

**Author contributions**

B.D. and W.Y. designed and fabricated the devices. W.Y. participated in initial part of measurements and fabrications. B.D. performed the measurements, with help of W.Y. in the initial part and H.K.K. in the later part. B.D., H.K.K. and R.A.M. participated in understanding the data, with throughout guidance of M.H. Y.O., A.S. and D.M. worked on the theoretical aspects. V.U. developed and grew the two-dimensional heterostructures.

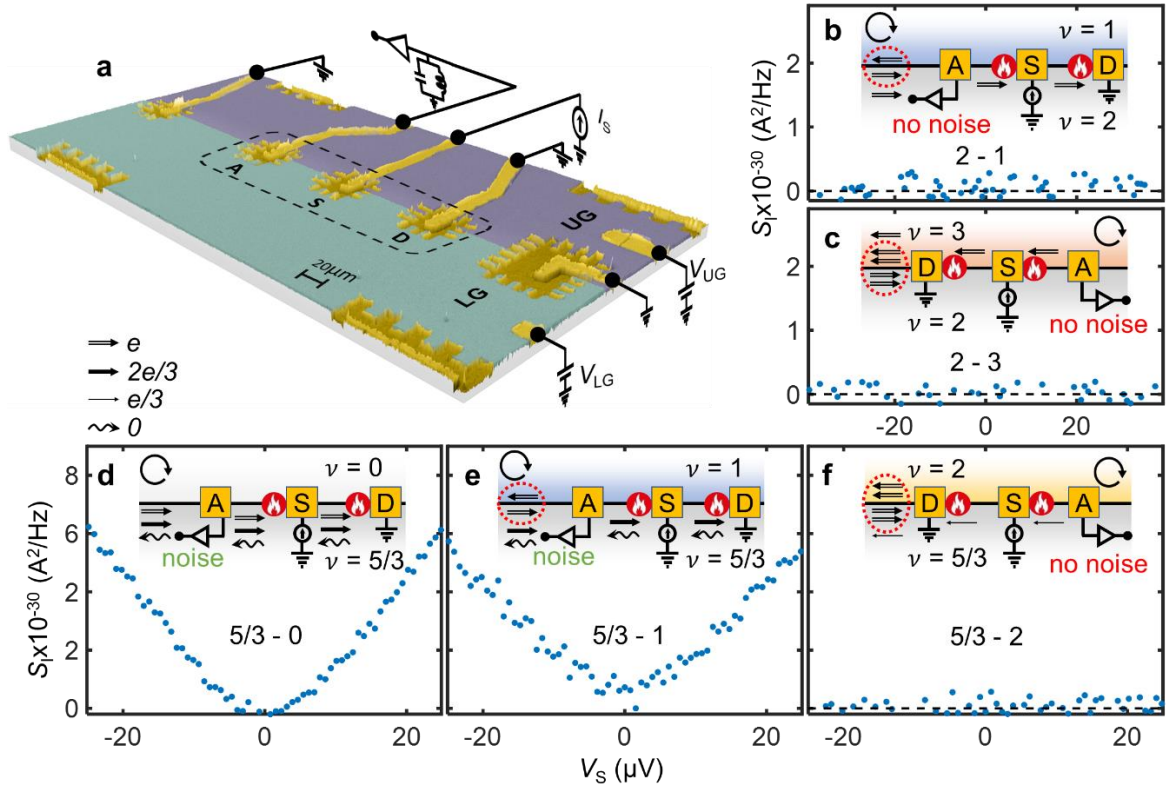

**Figure 1 | Experimental setup to create and probe the interface between different states. a,** False colors SEM image of a typical device. Ohmic contacts (shown in yellow). Lower-gate (LG, shown in light green) and upper-gate (UG, shown in purple). The 2DEG is buried 200nm below the surface (for detail structure refer to ED Sec. A). Gates' voltage, $V_{LG}$ and $V_{UG}$, control the density. The interface between the two planes hosts interface modes. Ohmic contacts at the edge probe the bulks' filling-factor. Ohmic contacts at the interface probe the interface modes. The heart of the device, highlighted by the dashed-box, contains the Source contact, S, placed at the same distance from the amplifier contact, A, and the Drain contact, D, at different distances S-A: $L$=28, 38, 48, and 58μm. 'Blocking contacts' (not shown) avoid noise arriving at A from secondary hotspots (say, in the Drain). **b, c,** Interfaces of integer states ν=2 with ν=1 & ν=3. Counter-propagating integer modes at the interface are gapped out, leaving a single integer mode at the interface, DS at 2-1 interface and US for 2-3 interface. Hotspots (shown by 'red fires') are shown on the US side (2-1) and the DS side (2-3) of the source. Noise is not expected. **d, e, f,** Interfacing the ν=5/3 state. Symbolic arrows stand for four different types of edge modes. Left-to-right symbols: non-equilibrated to equilibrated modes. The 5/3-0 and the 5/3-1 interfaces, presented in the particle-like picture: DS integer 1 and 2/3 modes and US neutral mode. The 5/3-2 presented in a hole-like picture: two DS integers 1 modes and an US 1/3 mode. The US neutral mode leads to noise in **d & e.** In **f,** a single US 1/3 mode remains, without any observed.

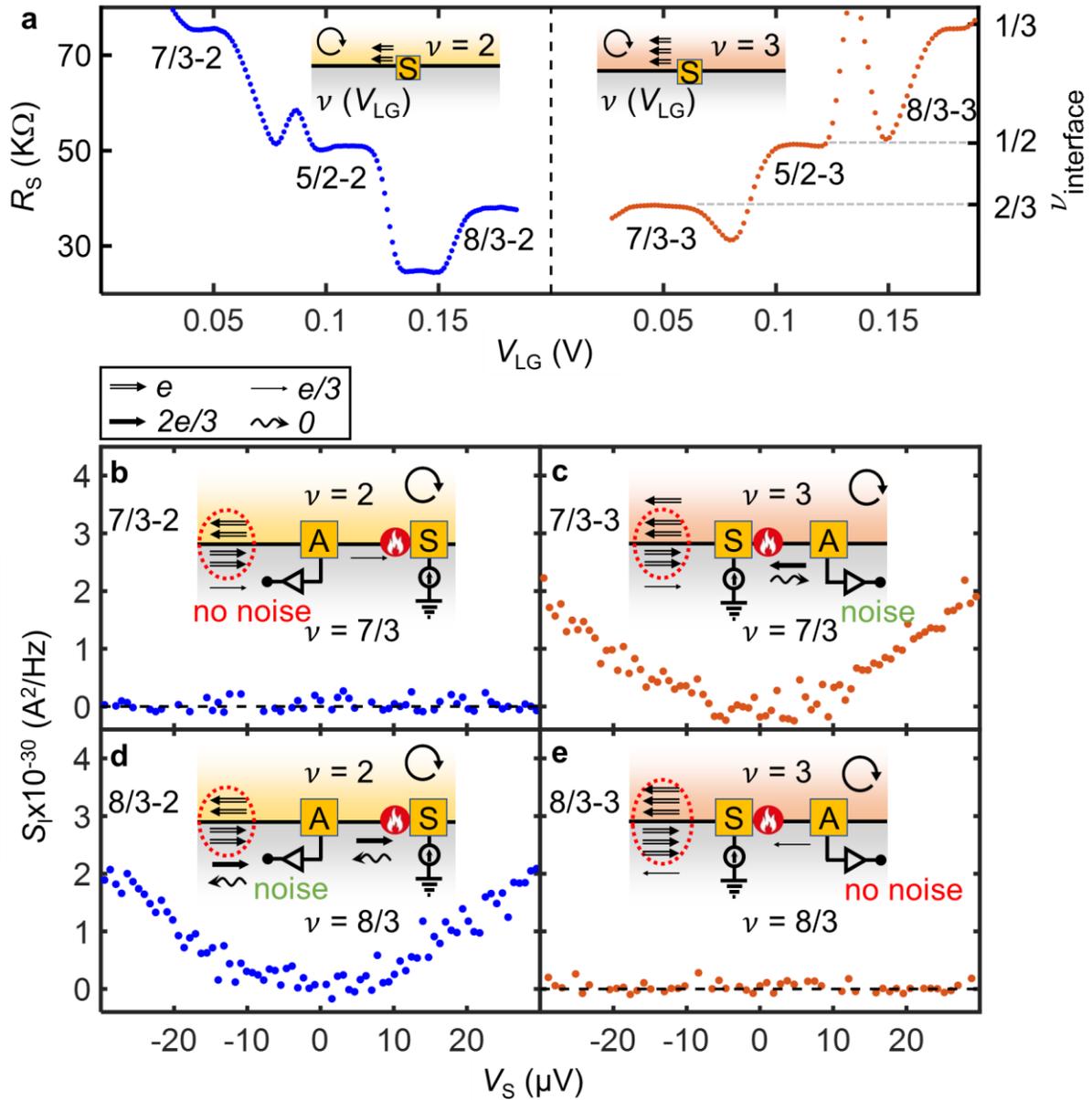

**Figure 2 | Interfacing Abelian fractional states in second Landau level. a**, Two-terminal resistance measured at the interface between ν=2 (or 3) and ν=7/3, 5/2 and 8/3. Left (right) inset: Upper plane is fixed at ν=2 (ν=3), while the lower-plane is swept from ν=7/3 to ν=8/3. Clear quantized plateaus corresponding to ν=1/3, 1/2 and 2/3 (left) and ν=2/3, 1/2, and 1/3 (right), accurate to ~1%, are observed. Peaks and valleys in between plateaus are due to re-entrant filling-factors. **b**, Interface between ν=7/3 and ν=2. The two integers are gapped, leaving the DS 1/3 charge mode. No noise observed. **c**, Interface between ν=7/3 & ν=3. Two integers gapped, leaving one US integer and a downstream 1/3. The equilibration of these two counter-propagating charge modes gives rise to an upstream ν=2/3 charge mode and a downstream neutral mode, accompanied by noise. **d**, Interface between ν=8/3 and ν=2. Two integers are gapped, leaving a DS ν=2/3 charge mode accompanied by an US neutral mode. US noise observed. **e**,

Interface between ν=8/3 and ν=3. The equilibration between the integers at the interface leaves a single US charge mode ν=1/3, with no noise.

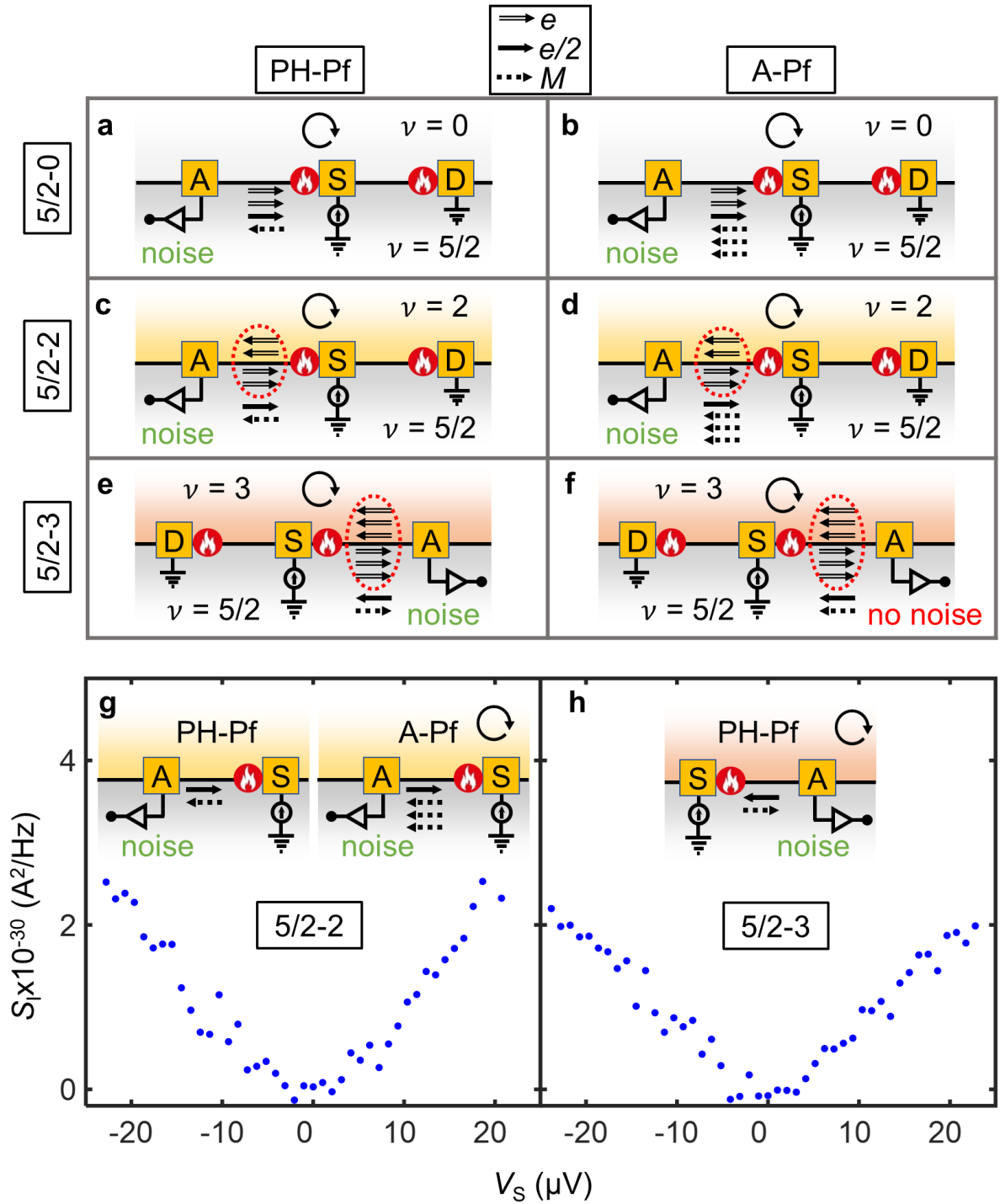

**Figure 3 Interfacing the ν=5/2 states with ν=2 and ν=3.** Comparison of edge-structures for the PH-Pf and the A-Pf orders with ν=0,2,3. **a**, **b**, Edge structure of ν=5/2 state at the edge of sample presented in 'particle-like' presentation. As US noise is expected for both orders, this measurement cannot distinguish between the two orders. **c**, **d**, Interfacing 5/2-2 for both the topological orders. US noise is expected for both orders. **e**, **f**, Interfacing 5/2-3 for both the orders, presented for convenience, in the hole-like picture. The three integer modes at the interface are gapped, leaving at the interface: PH-Pf - US 1/2 charge mode and a DS neutral Majorana mode; A-Pf - co-propagating US 1/2 charge and neutral Majorana modes. DS noise is expected for the PH-Pf order, while no noise is expected for A-Pf order. **g**, Measured US noise at the interface of 5/2-2, consistent with both of the competing orders. **h**, Measured DS noise at the 5/2-3 interface, expected only for the PH-Pf order.

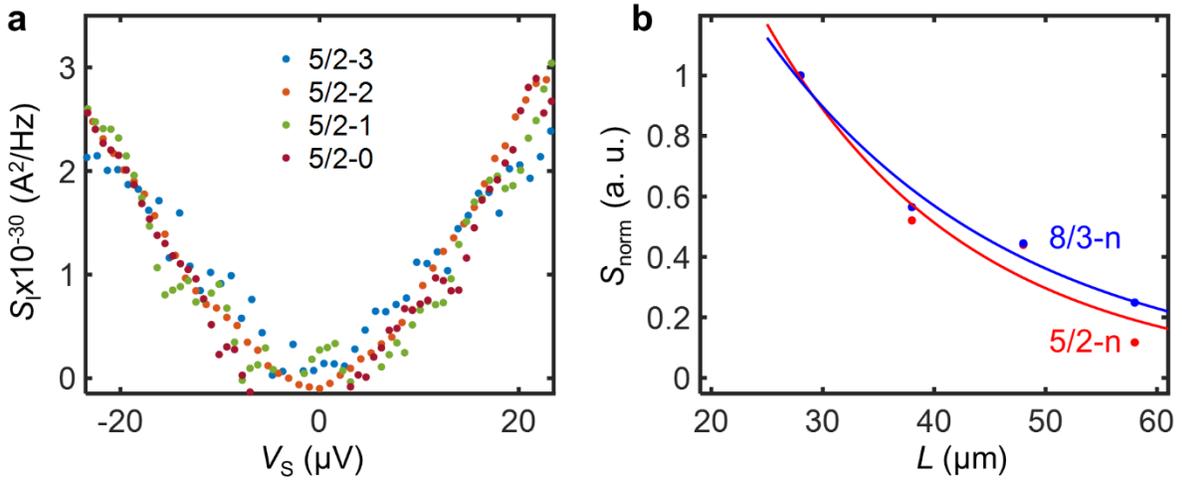

**Figure 4 | Dependence of the noise due to neutral modes on the number of integer modes and the propagation length. a,** Noise in a few interfacing conditions, 5/2-*n* with *n*=0, 1, 2, & 3, measured at 10mK and for 28μm propagation length. The measured US noise (and DS noise for 5/2-3) is *n* independent; indicating that the integer modes do not play a role in the excitation of the neutral modes. A similar observation is found also for 8/3-*n*, with *n*=0, 1, 2 (see also Fig. ED16b). **b**, The measured noise as a function of distance (normalized to the noise at 28μm). The solid lines are qualitative fits, drawn as a guide to eye. Since the noise amplitude does not depend on the number of integer modes (as seen in **a**), the data for ν=5/2 and ν=8/3 are annotated as 5/2-*n* and 8/3-*n*. The noise decay-length is qualitatively similar for both 5/2-*n* and 8/3-*n*.

# Extended Data

## Novel method distinguishing between competing topological orders

We provide more information on the following: The 2DEG structure design; Sample's fabrication steps; The gapping of integer modes; Interfacing of the 2/3 state; Equilibrated vs. un-equilibrated picture of the 5/2 state; The need of interfacing the 5/2 state with integers; Interfacing eight (out of nine) possible orders of the 5/2 state; Edge reconstruction of the bulk A-Pf order; Extended measurements at 28mK, 21mK and 10mK, for 28, 38, 48 and 58µm length, with a summary table; Length dependence of the noise at different 2/3-interfaces.

## A. 2DEG structure design

The heterostructures used in these experiments were designed to achieve a compromise between the robustness of the 5/2 fractional state and the ability to operate structure it with surface gates. It is well established that a quantization of the 5/2 state is governed by the long-range spatial potential landscape formed by the ionized donors. Smoothing this landscape is customarily achieved by excessive doping, either in 'short period super-lattice'[1], or in a low Al mole-fraction AlGaAs[2]. These doping schemes provide a significant reduction in the long-range potential fluctuations via spatial correlations between ionized donors, exhibiting, at the same time, negligible lateral parasitic conductance at low temperatures. Unfortunately, weakly localized electrons both thwart the stable operation of the surface gates and completely inhibit applying positive gate bias, which is essential in the present experiment. In order to solve the problem, we use the so-called 'inverted' 2DEG structure, where an accurate quantization of the 5/2 state is enabled by excess δ-doping in a layer of $Al_xGa_{1-x}As$ (x=0.23) placed 65nm (or 70nm) *below* the QW (Fig. ED1b). The surface potential is compensated using a uniformly doped layer made of $Al_{0.37}Ga_{0.63}As$:Si, located far away from the QW, thus minimizing the impact of its long-range random potential fluctuation on the 2DEG. Such high Al mole fraction doped layers are conventionally used between the gates and the 2DEG in most devices since the electrons freeze at T~100K in the DX-Si centers. The electron density in as-grown samples at 300mK is ~ $2.2 \times 10^{11} cm^{-2}$, and the mobility is $15 \times 10^6 cm^2/Vs$, even though the 2DEG wave function is asymmetrically shifted towards the bottom AlGaAs-GaAs interface - usually considered to be of a lower quality than the top one interface of the quantum well.

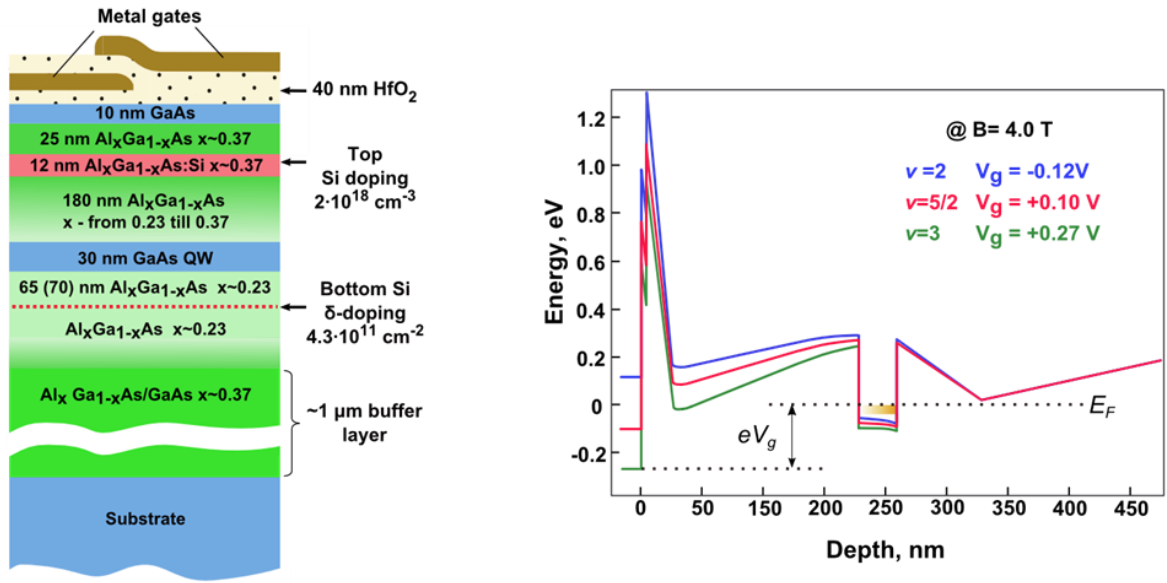

**Extended Data Figure 1| MBE design of the heterostructure. a.** Cross-section of the MBE growth (not in scale). **b.** Conduction band at three different gate voltages corresponding to three different filling factors $\nu$=2, 5/2, and 3.

## B. Sample fabrication process

We start with a 250 X 800μm MESA, made by wet etching in $H_2O_2:H_3PO_4:H_2O$=1:1:50 solution for 2 mins. A few ohmic contacts are deposited at the edge and at the middle of the MESA; annealed at 450°C for 2 mins. The whole sample is then coated with 25nm of $HfO_2$ layer via ALD. The half of the MESA is then patterned, followed by a deposition of 20nm Ti/Au metal thin film, which acts as the upper-gate (UG). In the next step, we coat the sample with another 15nm of $HfO_2$ layer, followed by another metallic plane covering the other half of the MESA, acting as the lower gate (LG). In the final step, the middle-ohmic contacts are connected to pads by thick gold lines; passing over the $HfO_2$ covered UG side, such that they do not short to gates. The upper half-plane is generally tuned to the integer regime.

## C. Interfacing integer edge-modes

To understand the new method used in our experiment, consider a region with a filling factor $v$=2, having two integer modes flowing at the edge, as shown on the left side of panels **a** and **b** of Fig. ED2. Interfacing this region with another region with filling factor $\nu$=1 (panel **a**) or $\nu$=3 (panel **b**), leads to interface modes flowing in opposite directions. Pairs of identical counter-propagating edge modes gap each other, leaving at the interface only the ungapped modes (Figs. ED2c & ED2d). We define the chirality of the interface modes according to the chirality of the side being probed, i.e., $\nu$=2 side. Hence, it is downstream (DS) in Fig. ED2c and upstream (US) in Fig. ED2d. The Landau levels (LLs) that correspond to these two cases are depicted in Figs. Ed2e & ED2f. Current at the edge flows where the LLs crosses the Fermi energy. The opposite slopes in the figures indicate that interface currents flow in opposite directions.

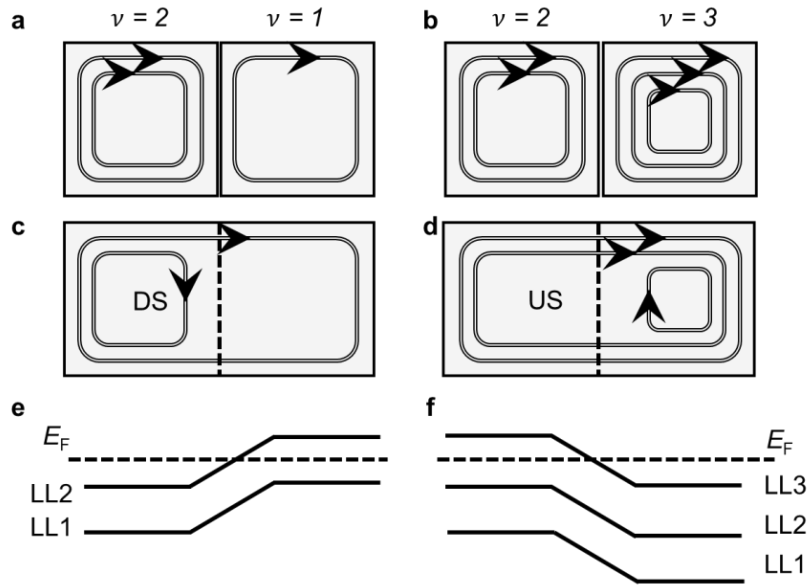

**Extended Data Figure 2 | a-d.** Interface two regions of different fillings, resulting in a single mode. **e-f.** The energy diagram of the LLs.

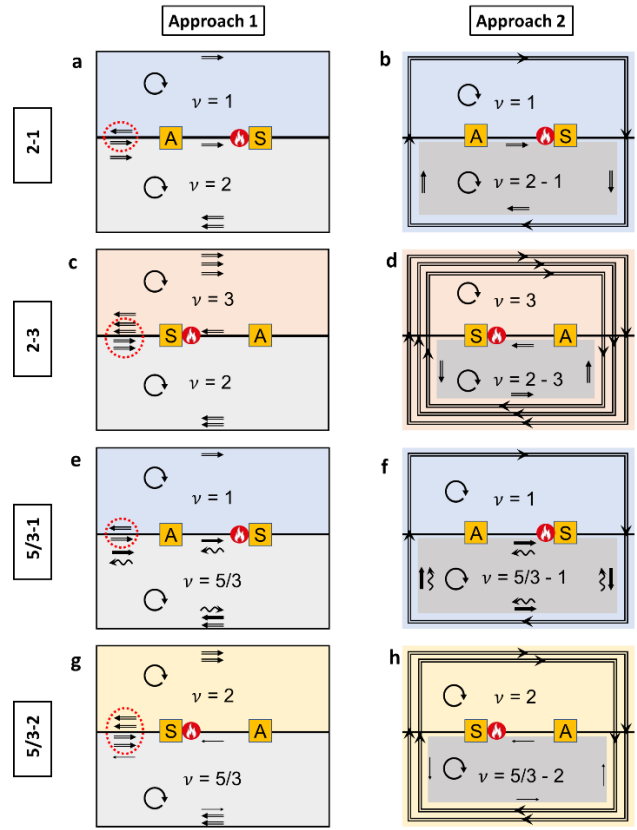

**Extended Data Figure 3 |Two equivalent approaches to understand the interface. a-b.** Interface between ν=2 and ν=1: Approach 1 - two regions of filling factor ν=2 and ν=1, are brought together and, as a result of equilibration, a single DS interface charge mode is formed. Approach 2 - one integer mode is encircling the MESA and a single integer mode remains at the interface (this assumes full mode equilibration at the interface). **c-d.** Interface between ν=2 and ν=3: Approach 1, two regions of filling ν=2 and ν=3 are brought together and equilibrate to a resultant US integer mode at the interface. Approach 2 - two integer modes gap each other at the interface, and thus circulate along the MESA edge, with an US integer mode at the interface. **e-f.** Interface between 5/3 and 1: Approach 1 - two separate half-planes of filling 5/3 and 1 equilibrate to a DS 2/3, and an US neutral at the interface. Approach 2 - one common integer mode moves along the edge of the full-plane, with a DS 2/3 and an US neutral at the interface. **g-h.** Interface between 5/3 and 2: Approach 1 -  two separate half-planes of 5/3 and 2 filling brought together, with an US 1/3 charge mode at the interface. Approach 2 - two integer modes move along the MESA edge, with US 1/3 charge at the interface.

The interfaceing method can be visualized in two equivalent descriptions. Figure ED3 describes these two approaches for the interface of integer-integer: '2-1', '2-3' and integer-fraction in first LL: '5/3-1', '5/3-2'. The first approach considers the two separate half-planes with their own filling-factors. With close proximity of the two half-planes at the interface, the integer modes gap out, leaving the resultant interface modes at the interface. The second approach is equivalent to the first one, but it considers that the integer modes are fully gapped and thus moving along the edge of the mesa, thus do not exist at the interface. Although these two approaches are equivalent, the first one is easy to visualize for the cases of complex interfacing states, e.g., interfacing of '4/3-2/3' leading to a resultant $\nu$=2/3 mode consisting of two copropagating 1/3 modes (without neutral mode). Therefore, in the main text, we adopted the first approach to describe our experimental observations.

### D. Application of the interfacing method to $\nu$=2/3 state

The charge unequilibrated hole-conjugate $\nu$=2/3 state in the lowest LL supports a DS integer charge mode $\nu$=1 and a US fractional $\nu$=1/3[3,4]. With charge equilibration, a $\nu$=2/3 charge and an US neutral modes are formed[5-8]. This process is depicted in Fig. ED 4a & 4b for the interface of $\nu$=2/3 and 0. The equilibration process between the two counter-propagating modes near the Source leads to a hotspot at the back of the Source (and at the front of drain, D). The neutral mode, excited by the hotspot, carries its energy US to produce noise at the Amplifier contact, as indeed is found. However, when $\nu$=2/3 is interfaced with $\nu$=1, it leads to US $\nu$=1/3 charge mode at the interface. No neutral mode exists

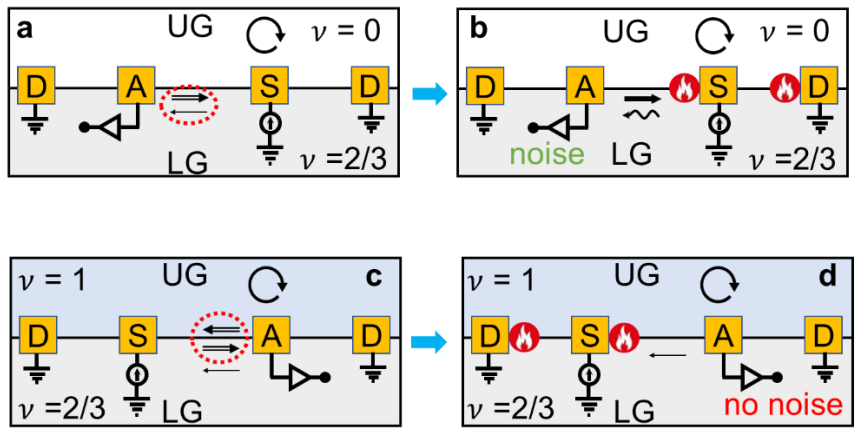

**Extended Data Figure 4 |** Interface 2/3-0 and 2/3-1. Upper-panel (**a**, **b**) - equilibration between 2/3 and 0 produces US noise. Lower-panel (**c**, **d**), Gapping the integer mode at the interface 2/3-1, leaves US 1/3 mode. Double-line arrow denote integer modes, single thick-line arrow for the 2/3 mode, single thin-line arrows for the 1/3 mode, and wiggling arrow for a neutral modes.

## E.  $\nu$ =5/2 state: equilibrated *vs.* un-equilibrated picture

The primary two candidates of the topological orders of the $\nu$=5/2 state can be presented in the following two pictures: ***i***) un-equilibrated, or the so-called 'hole-picture', and ***ii***) equilibrated, the 'particle-picture' (Fig. ED5). In the un-equilibrated picture, the PH-Pf order hosts three DS integer modes, a DS Majorana mode, and US ½ charge mode. The A-Pf order hosts three DS integer modes, US ½ charge mode and a Majorana mode. In the equilibrated picture, both of the orders host DS two integer modes and a ½ charge modes, and both host upstream neutrals (single Majorana for PH-Pf and three Majorana modes for A-Pf) modes. In experiments on the edge, i.e., at the 5/2-0 interface, we always find an US neutral, thus excluded from the start the Pf order; however, this cannot distinguish between the two proposed orders.

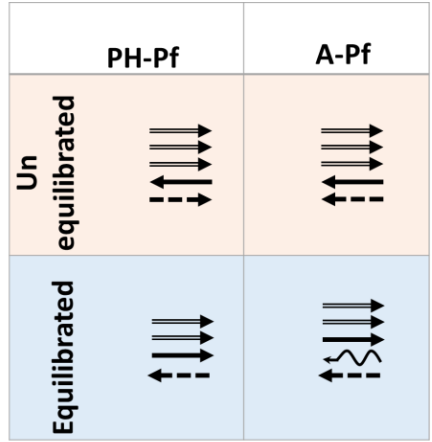

**Extended Data Figure 5 |** Two possible pictures of the $\nu$=5/2 state for PH-Pf and A-Pf orders. Double-line arrow -- integer, single-line arrow -- 1/2 charge mode, dashed-arrow -- Majorana mode, and wiggling-arrow -- neutral mode.

## F.  The need of interfacing $\nu$=5/2 with integers

The proposed way to prevent inter-mode equilibration between the counter-propagating fractional and integer charge modes. This can be done by equilibrating the 5/2 state with integers modes, thus gapping them out.

*5/2-2 interface***:** Figure ED6 shows the equilibration process at the interface of 5/2-2, for both PH-Pf and A-Pf orders.

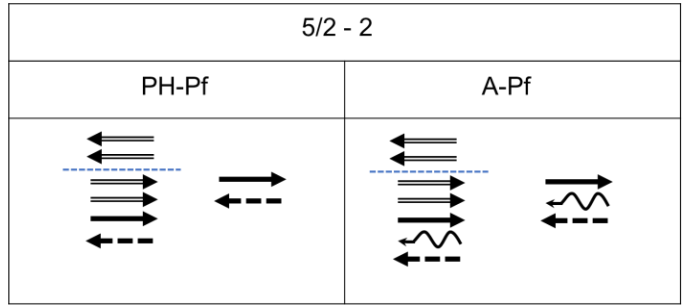

**Extended Data Figure 6 |** Resultant modes structure at the interface 5/2-2: with both having a DS ½ charge mode and US Majorana neutral modes.

As we can see, both orders possess a DS ½ charge mode and US Majorana modes (one for PH-Pf and one neutral with a Majorana for A-Pf). Note, a DS charge mode, which is needed to ignite the required hotspot exists in both cases. So, interfacing 5/2 with 2 cannot distinguish between PH-Pf and A-Pf.

*5/2-3 interface:* Figure ED7 described the equilibration process at 5/2-3. Both orders contain an US ½ charge mode, but the Majorana mode is moving in the opposite direction. The interface modes are now distinguishable. Noise in the DS direction supports the PH-Pf order.

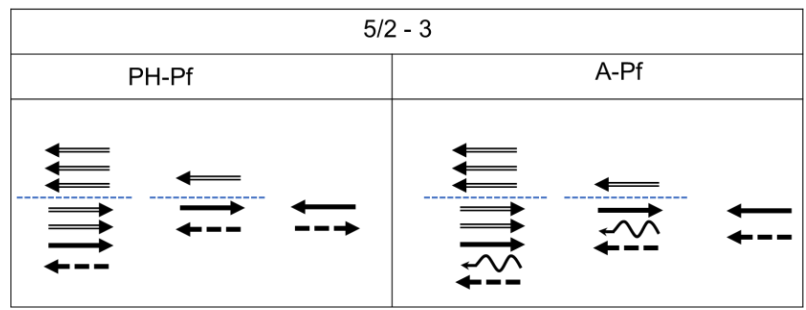

**Extended Data Figure 7 |** Interface edge structures of 5/2 & 3 are distinct between PH-Pf and A-Pf, with PH-Pf supports counter-propagating ½ charge and Majorana mode, while A-Pf supports co-propagating US ½ charge and Majorana mode.

## G. The interfacing-method with all nine-proposed orders of $\nu$=5/2

We considered nine proposed topological orders for the $\nu$=5/2 state (Fig. ED8): four abelian orders('331', 'K = 8', '113', 'anti-331') and five non-abelian orders ('SU(2)$_2$', 'Pfaffian', 'PH-Pfaffian', 'anti-Pfaffian', 'anti-SU(2)$_2$')[9-11].

Simple conductance and noise measurements cannot determine the correct order of the state. In the main text, we describe our experimental results, considering the two most favourable non-abelian orders: PH-Pf and A-Pf. Here, we show that the interfacing method with the nine possible orders of $\nu$=5/2 state. In the absence of edge reconstruction, only the PH-Pf agrees with our measurements.

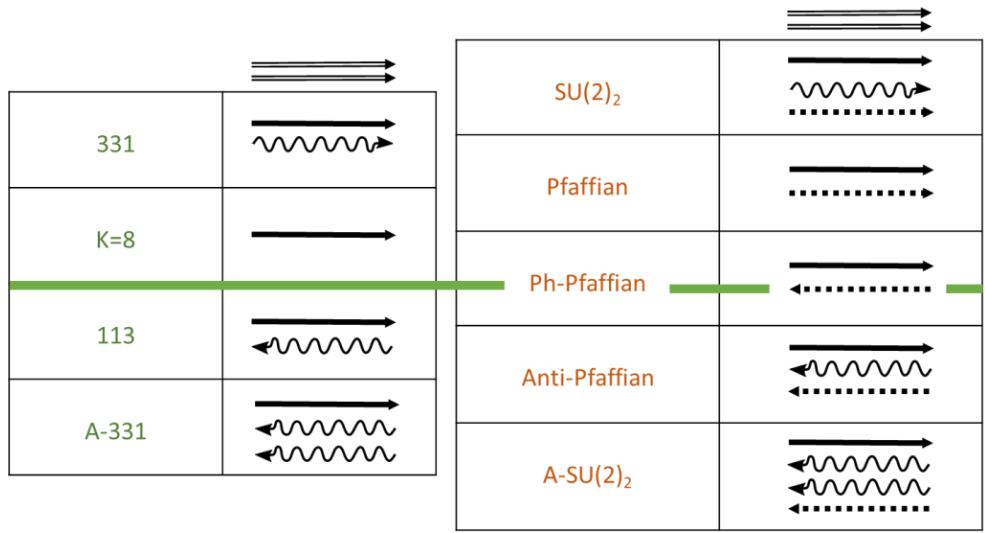

**Extended Data Figure 8 |** Nine orders of the 5/2 state: Left-column shows the 4 Abelian orders, while in the right column the 5 non-Abelian orders are plotted. The green line stands for the boundary between particle-like and hole-like orders, with PH-Pfaffian having particle-hole symmetry.

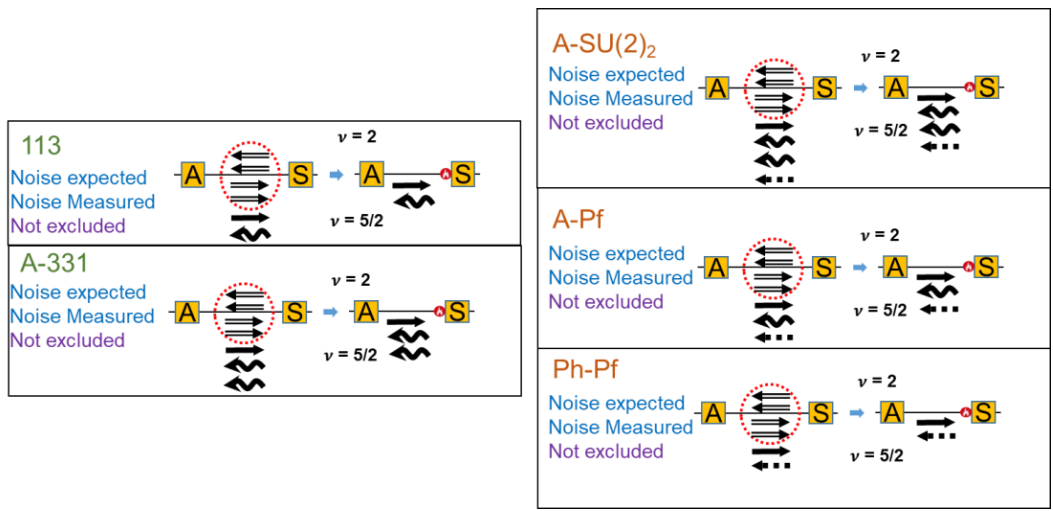

**Extended Data Figure 9 |** Interfacing with $\nu$=2 cannot exclude any of the possible

*5/2-0 interface:* The observed US noise at 5/2-0 interface is inconsistent with the edge structure of, '331' and 'K = 8', from abelian orders, and the 'SU(2)$_2$' and the 'Pfaffian' from non-abelian orders, thus rulling them out.

We now look at: '113', 'A-331' abelian orders and at: 'PH-Pf', 'A-Pf', and 'A-SU(2)$_2$' non-abelian orders.

*5/2-2 interface:* Observing upstream noise cannot exclude any of the 5-possible orders (Fig. ED9).

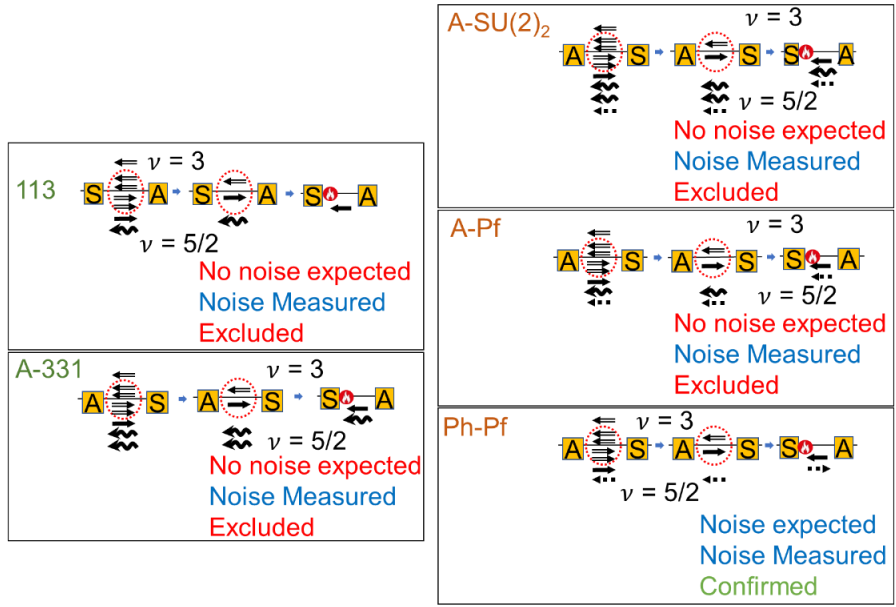

**Extended Data Figure 10 |** Interface with $\nu$=3 shows that only PH-Pf order can explain out data.

*5/2-3 interface:* As seen in Fig. ED10, only the PH-Pf supports a measurement of noise at the amplifier in DS, matching with our experimental observations.

## H. Edge reconstruction of bulk A-Pf and Pf at the interface of *v*=3

One may consider the possibility that the bulk topological order is the A-Pf; however, with the interface is edge-reconstructed to have a strip of a Pf order *only* when it is interfaced with $\nu$=3. In this scenario, there are now two interfaces: 3–5/2 (Pf) and 5/2 (Pf)–5/2 (A-Pf) - separated by the reconstructed Pf-order strip. With an additional assumption that the width of the strip is large enough to make the coupling between the modes at the two interfaces weak, the interface 3 –5/2 (Pf) mode, has an excited DS neutral modes, which may lead to an observed noise (see Fig. ED11).

While we cannot entirely rule out this scenario, we deem it improbable for several reasons. Firstly, since the measured noise does not depend on the interfaced integers (Fig. 4a in the text), it is hard to believe that the DS noise, generated by the interface of the Pf strip with $\nu$=3, will be identical to the noise of the A-Pf at different interfacing conditions without the reconstruction. Secondly, the coupling across the narrow Pf-strip (being on the scale of the depth of the 2DEG below the gates, ~200nm); hence, changing the interface structure (with different integers, not ν=3) should have led to a different decay length of the noise, which we do not find.

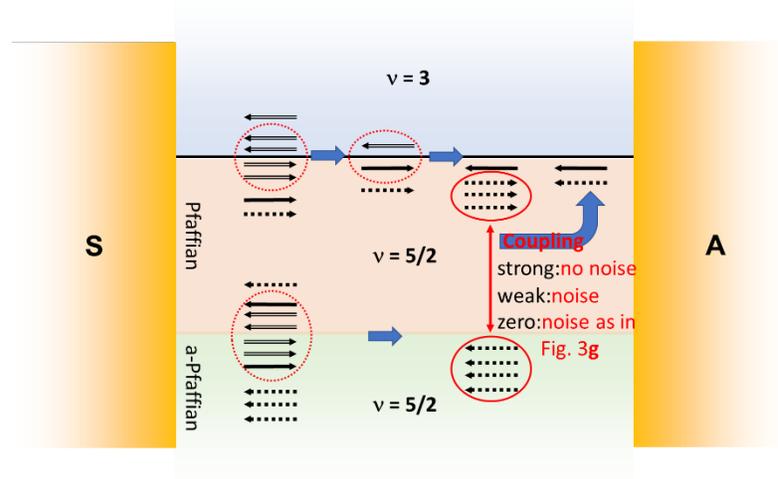

**Extended Data Figure 11** | Edge reconstruction at the edge of the sample. Zero inter-strip coupling: similar noise as in the experiment (Fig. 3g in main text); weak-medium coupling: smaller noise with increasing propagation length, being inconsistent with the experiment; strong coupling: no noise is inconsistent with the experiment.

## I. Extended measurements

In this section we give the detailed measurement data for the combinations of 3 different temperatures $T$=28, 21 and 10mK, and for 4 different Source-Amplifier propagation lengths, $L$=28, 38, 48, 58μm, for the interfaces with particle-states and hole-states in the lowest-LL, $1^{st}$-LL and $2^{nd}$-LL. The observed noise was found to decay exponentially with increasing temperature and propagation length from source to amplifier. Therefore, for higher bath temperatures, 21mK and 28mK, we considered noise measurement only at the smallest length 28μm. All the measurements are again summarized in Table 1,2.

### (i) $T$ = 28mK

Here, we present all the measurement data taken at 28mK and with a propagation length of 28μm. The data covers all the interfacing states in the $1^{st}$ and $2^{nd}$ LL. The observed noise satisfies the expected edge structures of the known states. As in the main text, we observed noise at the 5/2-3 interface, with a same conclusion as for the 10mK data in the main text.

**2nd LL: ν = 5/2, 8/3, 7/3 interface with ν = 2, 3:**

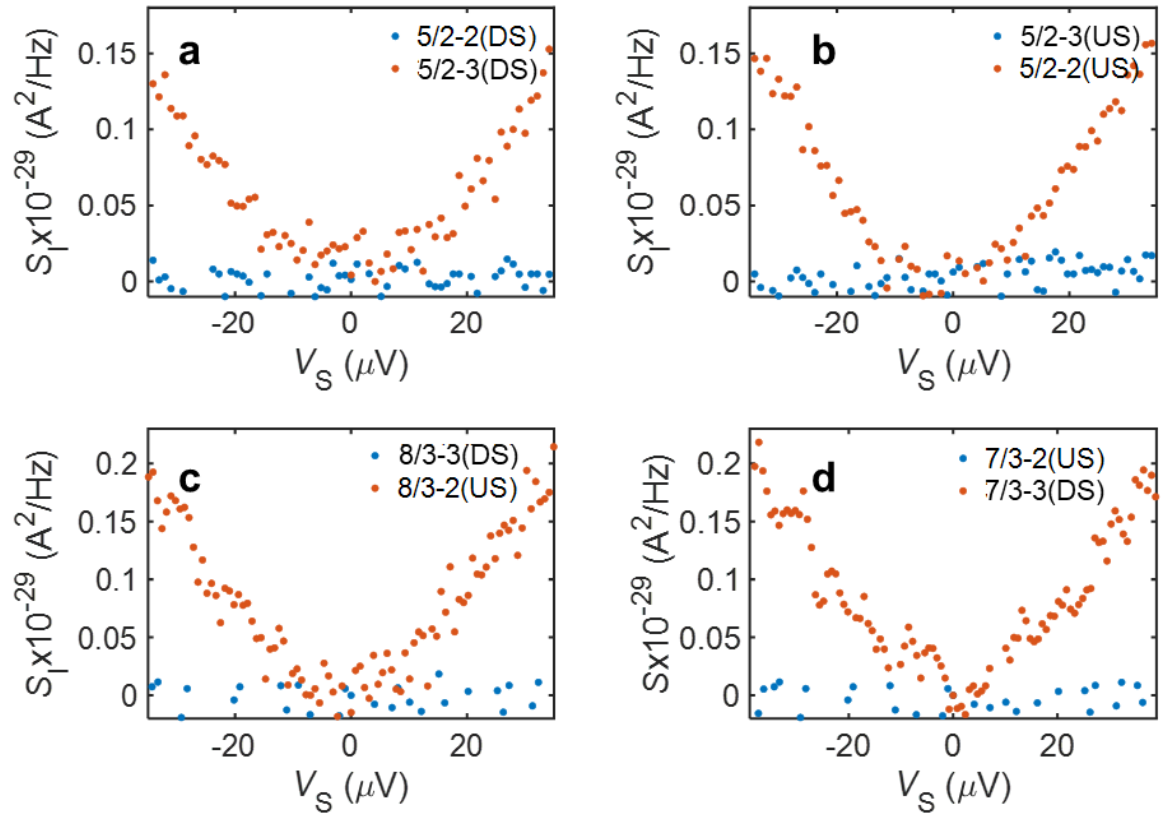

**Extended Data Figure 12 |** Noise measurement at 28mK, 28μm length for the interfaces of integer modes with the 2nd LL states. Noise at the 5/2-3 - again indicates the order of ν=5/2 to be PH-Pf.

**1st LL, Hole-conjugate states: ν = 1+2/3, 1+3/5, 1+4/7 interface with ν = 1 and 2:**

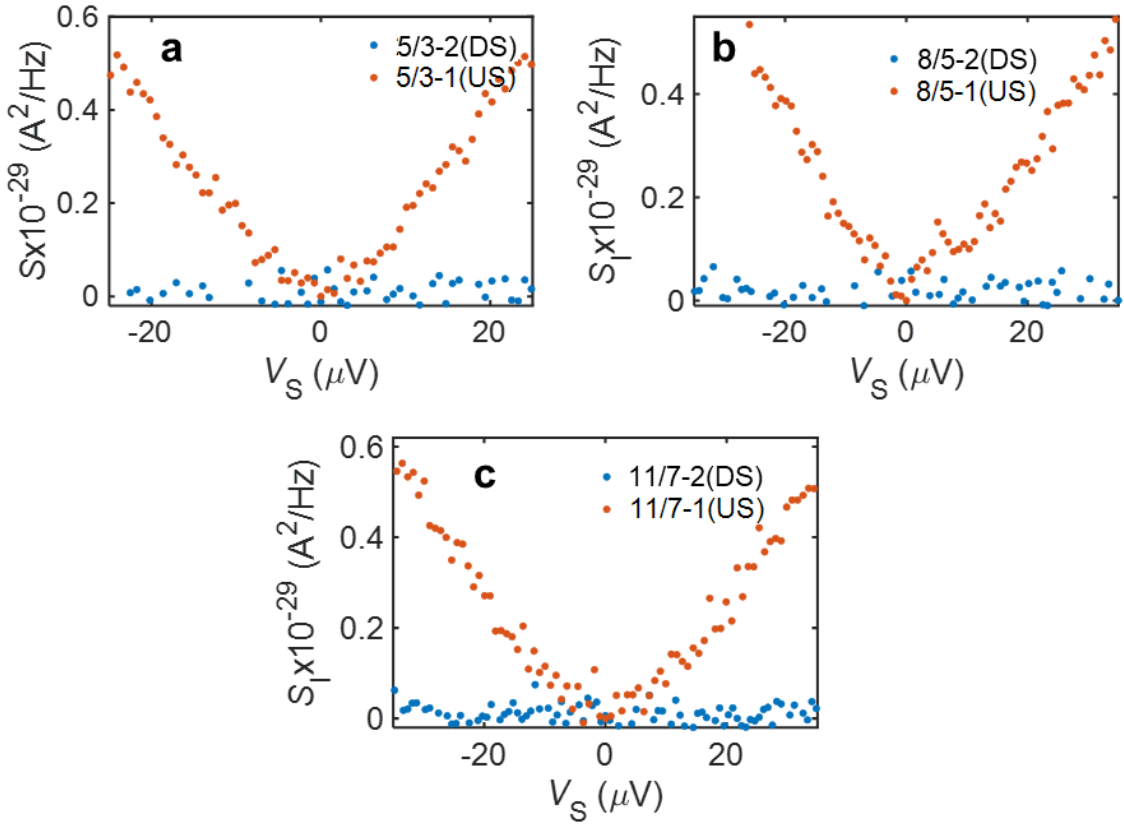

**Extended Data Figure 13 |** Noise at 28mK, 28mu length for the interfaces of 1+2/3, 1+3/5 and 1+4/7 with 1 and 2; noise is found for interfaces with 1, while no noise found for interfaces with 2, as expected.

**1st LL, Particle states: ν = 1+1/3, 1+2/5 interface with ν = 1:**

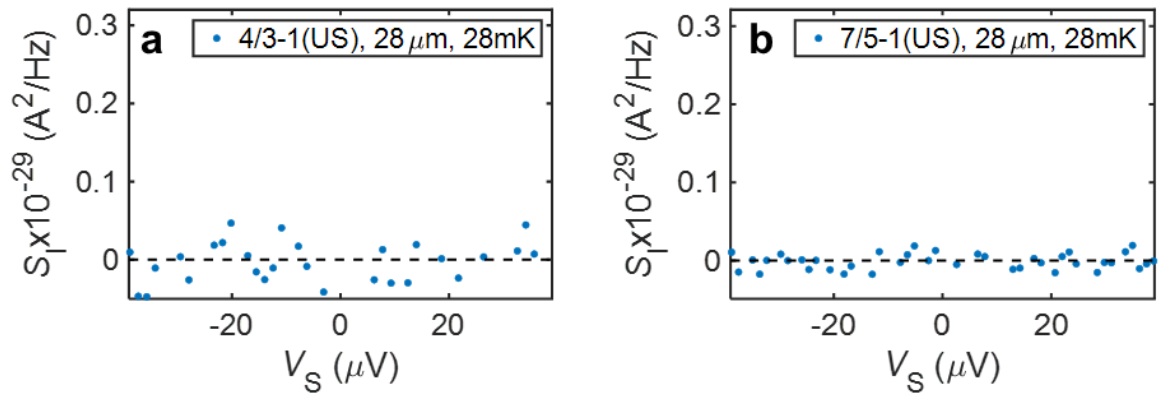

**Extended Data Figure 14 |** No noise observed at the interfaces of 1+1/3, 1+2/5 with 1, as expected from edge structure.

## (ii) $T$ = 21 mK

We present all the measurement data taken at 21mK and with a propagation length of 28μm. The observed noise satisfies the expected edge structures of the known states. As in the main text, we observed noise at 5/2-3 interface, again probed the order of the ν=5/2 state to be a PH-Pf.

**$2^{nd}$ LL: ν = 5/2, 8/3, 7/3 interfaced with ν= 2, 3**

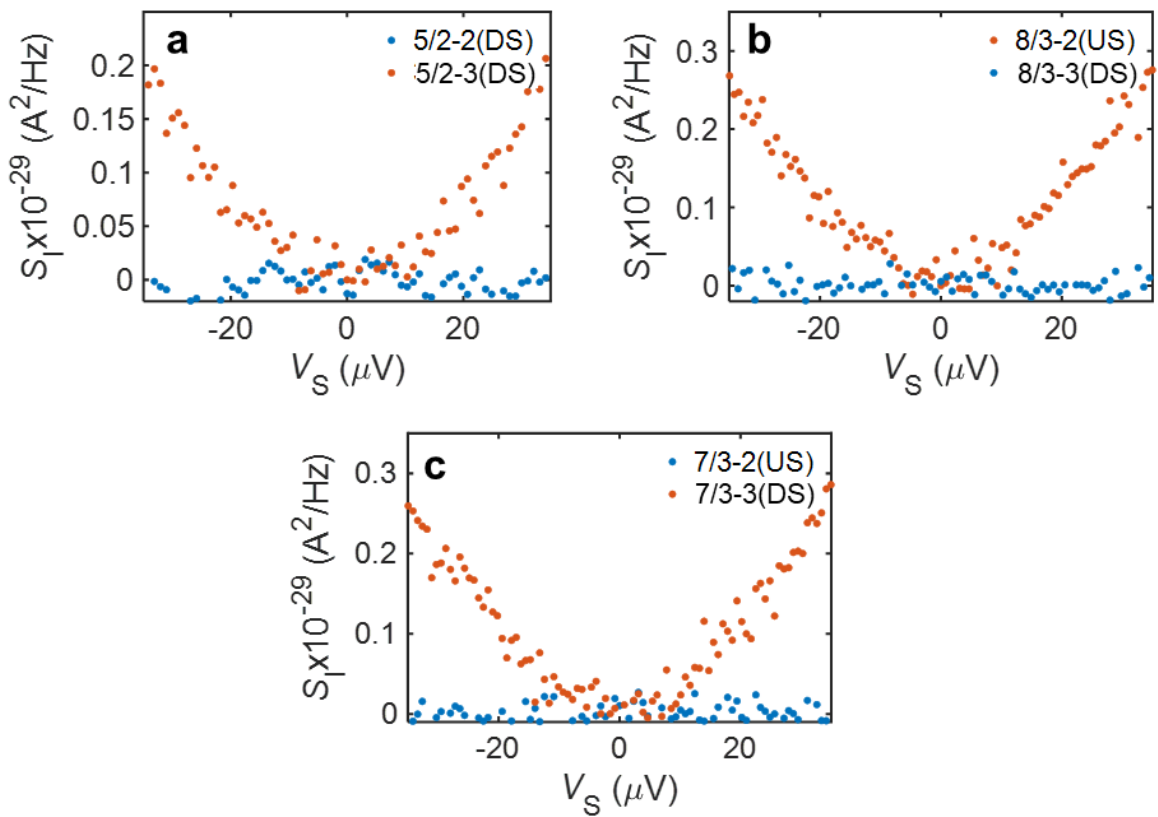

**Extended Data Figure 15 |** Noise at the interfaces of 5/2, 8/3, 7/3 with 2 and 3 at 21mK for 28μm length. The observed data matches with a PH-Pf order of the ν=5/2 state, as for the 10mK data in the main text.

### (iii) *T* = 10mK

In this subsection we present all the measurement data taken at 10mK and with a propagation length of 28, 38, 48 and 58μm. The data in the main text belongs to temperature 10mK, and propagation length of 28μm, with the additional data at 10mK are shown here. The conclusion from the observed noise at the 5/2-3 interface, at different lengths, remains the same as in the main text.

### *28 μm propagation length*

**2$^{nd}$ LL: ν = 5/2, 8/3 interface with ν = 2, 3:**

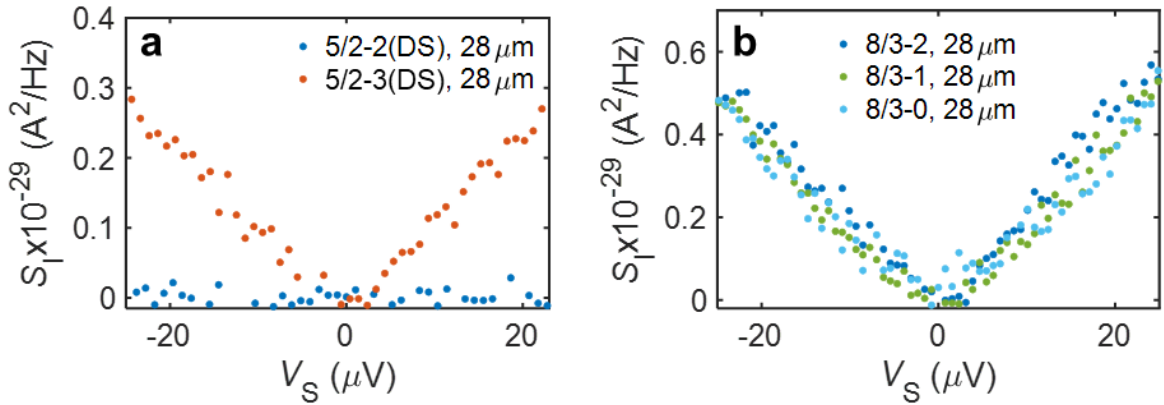

**Extended Data Figure 16 |** Noise at the interface of 5/2 with 2 and 3, and 8/3 with 2, 1 and 0 at 10mK for 28μm propagation length.

**Lowest LL, Hole-conjugate state: ν=2/3 interface with ν=0:**

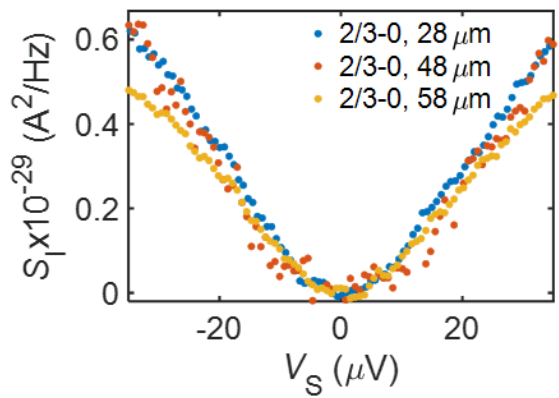

**Extended Data Figure 17 |** Noise at the interface of 2/3-0 at 10mK as function of length.

### *38 μm propagation length:*

**2nd LL: ν = 5/2, 8/3 interface with ν=2, 3:**

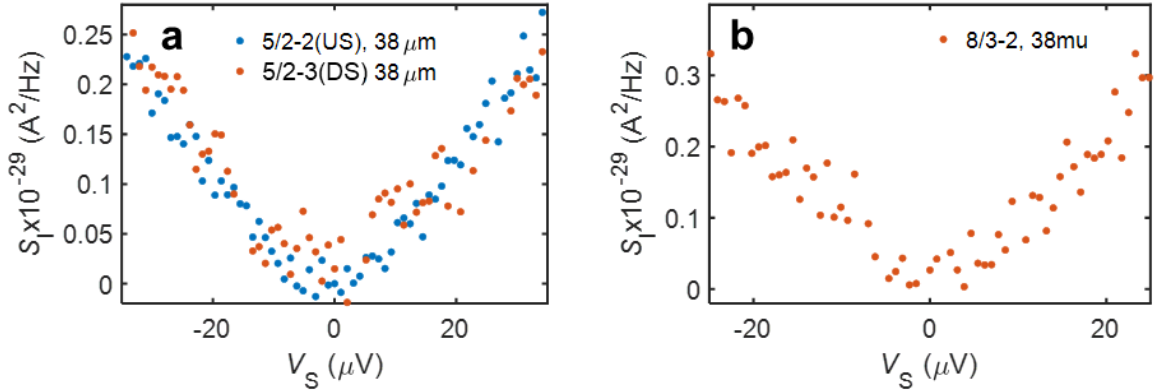

**Extended Data Figure 18 |** Noise at the interface of 5/2 with 2 and 3, and 8/3-2 at 10mK, for a propagation length of 38µm. The observed noise in 3-5/2 interface indicated PH-Pf order for the ν=5/2 state.

### *48 μm propagation length*

**2nd LL: ν = 5/2, 8/3, 7/3 interface with ν =0, 1, 2, 3:**

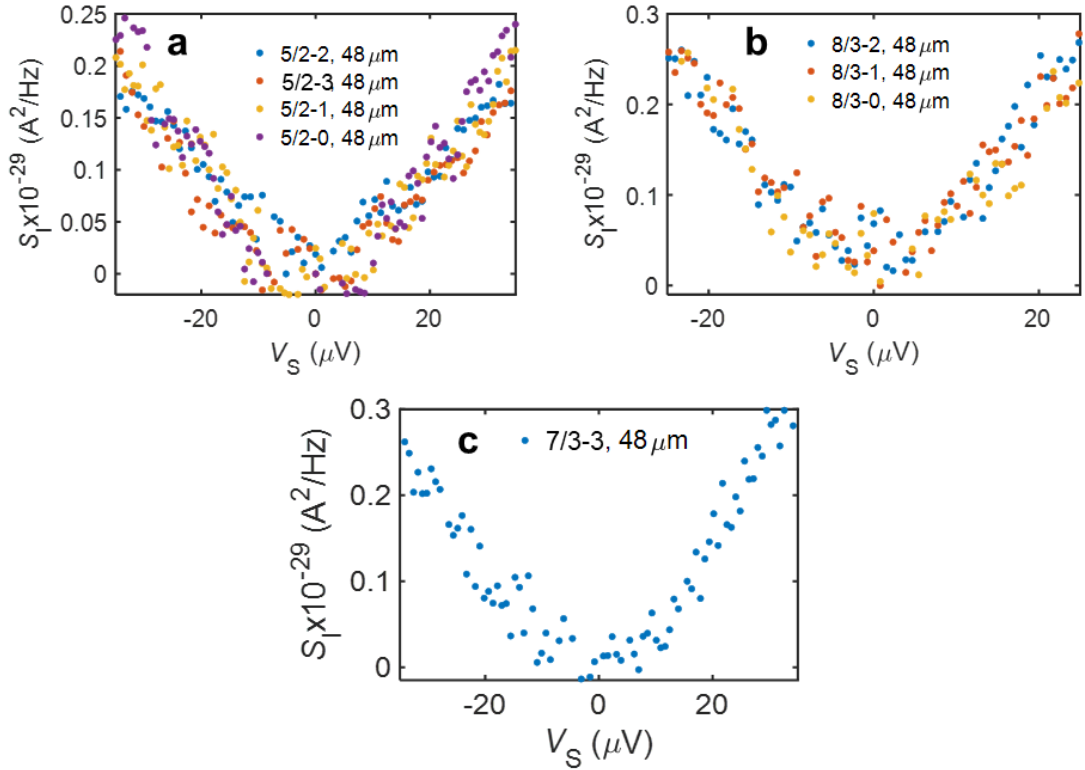

**Extended Data Figure 19 |** Noise at the interface of 5/2, 8/3, 7/3 with 2,3; 10mK, 48 µm, with a same conclusion as in the main text.

**1st LL, Hole-conjugate state: ν =1+2/3 interface with ν=1:**

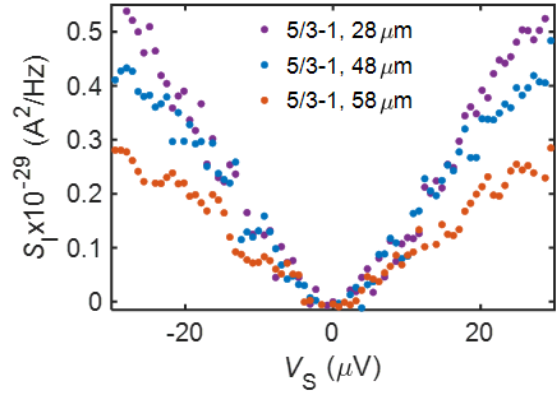

**Extended Data Figure 20 |** Noise measurement at the interface of 1+2/3 with 1 at 10mK at different lengths.

## 58 μm propagation length

**2nd LL: ν=5/, 8/3 interface with ν=0, 1, 2:**

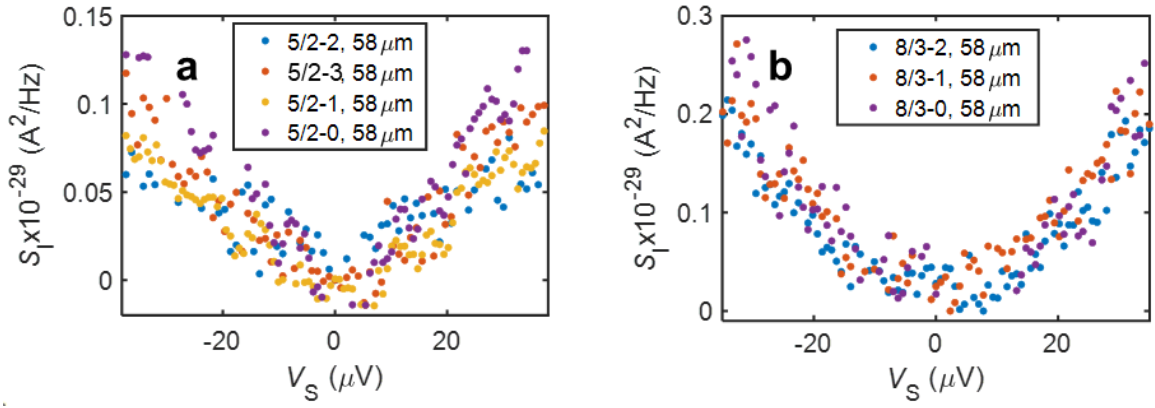

**Extended Data Figure 21 |** Noise measurement at the interface of 5/2, 8/3 with 0,1,2,3 at 10mK for a propagation length of 58μm.

## Summary Table

| $\nu_{LG}$ | 28 mK Length / $\nu_{interface}$ | 28 μm | 21 mK Length / $\nu_{interface}$ | 28 μm |
|---|---|---|---|---|
| 5/2 | 5/2-2 | Noise(US)/No noise(DS) | 5/2-2 | Noise(US) |
|  | 5/2-3 | No Noise(US)/ Noise(DS) | 3-5/2 | Noise(DS) |
|  | 5/2-1 | Noise | 5/2-1 | -- |
|  | 5/2-0 | Noise | 5/2-0 | -- |
| 8/3 | 8/3-2 | Noise | 8/3-2 | Noise |
|  | 8/3-3 | No noise | 3-8/3 | No Noise |
|  | 8/3-1 | Noise | 8/3-1 | -- |
|  | 8/3-0 | Noise | 8/3-0 | -- |
| 7/3 | 7/3-2 | No noise | 7/3-2 | No noise |
|  | 7/3-3 | Noise | 7/3-3 | Noise |
| 5/3 | 5/3-1 | Noise |  |  |
|  | 5/3-2 | No noise |  |  |
| 8/5 | 8/5-1 | Noise |  |  |
|  | 8/5-2 | No noise |  |  |
| 11/7 | 11/7-1 | Noise |  |  |
|  | 11/7-2 | No noise |  |  |
| 4/3 | 4/3-1 | No noise |  |  |
| 2/3 | 2/3-0 | Noise |  |  |

**Table 1**: Summary of all measurements at 28mK and 21mK; Amplifier's position (unless it is stated as US/DS) is in-line with the hotspot.

| $\nu_{LG}$ | Length / $\nu_{interface}$ | 28 μm | 38 μm | 48 μm | 58 μm |
|---|---|---|---|---|---|
| | | | **10 mK** | | |
| 5/2 | 5/2-2 | Noise(US)/No noise(DS) | Noise | Noise | Noise |
| | 5/2-3 | No Noise(US)/ Noise(DS) | Noise | Noise | Noise |
| | 5/2-1 | Noise | -- | Noise | Noise |
| | 5/2-0 | Noise | -- | Noise | Noise |
| 8/3 | 8/3-2 | Noise | Noise | Noise | Noise |
| | 8/3-3 | No noise | No noise | No noise | No noise |
| | 8/3-1 | Noise | -- | Noise | --- |
| | 8/3-0 | Noise | -- | Noise | --- |
| 7/3 | 7/3-2 | No noise | No noise | No noise | No noise |
| | 7/3-3 | Noise | Noise | Noise | Noise |
| 5/3 | 5/3-1 | Noise | -- | Noise | Noise |
| | 5/3-0 | Noise | -- | | |
| 2/3 | 2/3-0 | Noise | -- | Noise | Noise |

**Table 2:** Summary of all measurements at 10mK; Amplifier's position (unless it is stated as US/DS) is in-line with the hotspot.

## J. Noise decay length at different 2/3 interfaces

We also studied how the thermal equilibration length depends on the propagation length of the ν=2/3 mode created by interfacing fractional states in different LLs with integer modes; namely, at the interfaces of 8/3-2, 5/3-1 and 2/3-0 (Fig. ED22a). It is found that the noise due to the excited neutral modes decays faster for the 2/3-interface created with higher LLs. Below we describe three situations in some detail.

*Case 1, ν=2/3 interface created in the second LL, i.e., 8/3-2 interface*: We tune the LG side to ν=8/3 and UG to ν=2. At the interface, two integer modes are gapped out, allowing only a ν=2/3 charge mode and an US neutral mode, and in the bulk (ν=8/3) 2/3 mode is protected by two integer modes of the native state (Fig. ED22d).

*Case 2, ν=2/3 interface in the first LL, i.e., 5/3-1 interface*: Here we tune the LG side to ν=5/3 and UG side to ν=1. At the interface, one integer mode is gapped, with a resulting 2/3 charge mode and US neutral mode, and in this case the 2/3 mode in the bulk is protected by a single integer mode (Fig. ED22c).

*Case 3, ν=2/3 interface in lowest LL, i.e., 2/3-0:* In this case we tune the LG side to ν=2/3 and fully pinch the UG side to ν=0. Therefore, this is similar as the ν=2/3 state on the physical edge of the sample. In this case of native ν=2/3 state, with 2/3 charge mode and US neutral mode, both bulk and the interface has same edge structure and 2/3 mode is not protected by any integer modes (Fig. ED22b).

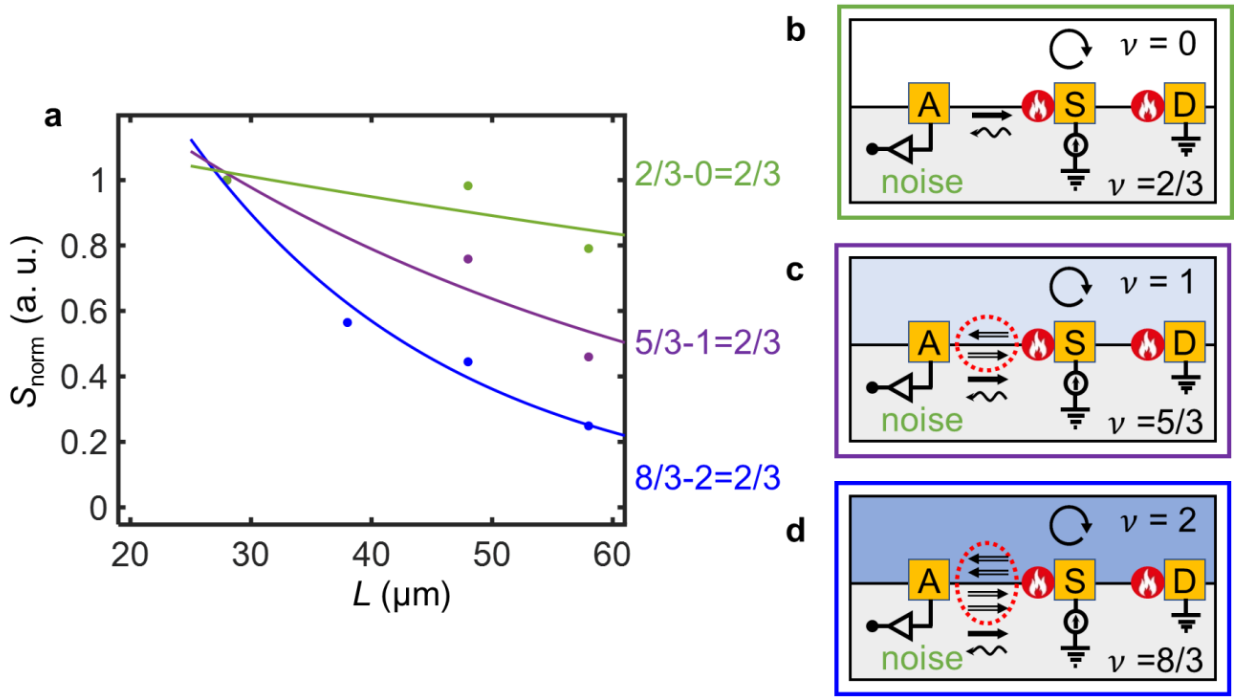

**Extended Data Figure 22 |** Length dependence of the noise at 10mK for a similar ν=2/3 interfaces, created differently by interfacing *8/3-2*, *5/3-1* and *2/3-0*. The decay length is found to decrease as the interface is formed by higher LLs.

The equilibration length $l_{eq} \sim (v_c + v_n)$, with $v_c$ the velocity of charge mode and $v_n$ the velocity of the neutral mode[12]. In an intuitive picture, the equilibration length increases when there is a large velocity mismatch between the charge and neutral modes, such that the inter-mode tunneling and hence the equilibration becomes difficult. The velocity of the charge mode $v_c \sim \varepsilon/B$, where $\varepsilon$ is the electric field and *B* the magnetic field. Therefore, with an increased the electric field (at constant *B*), $v_c$ increases and $l_{eq}$ increases, and hence a slow equilibration is expected.

In the above three cases, $v_c \sim \varepsilon/B$ is relatively constant (in our experiment), so one should expect to have similar $v_c$, and hence, a similar equilibration length. This experimental discrepancy may suggests that the integer modes in the bulk screen the electric field.